\documentstyle[12pt,psfig]{article}
\newcommand{\MEV}{~\mbox{MeV}}
\newcommand{\GEV}{~\mbox{GeV}}
\newcommand{\TEV}{~\mbox{TeV}}
\newcommand{\gsim}{ \mathop{}_{\textstyle \sim}^{\textstyle >} }
\newcommand{\lsim}{ \mathop{}_{\textstyle \sim}^{\textstyle <} }
\newcommand{\vev}[1]{ \left\langle {#1} \right\rangle }

\begin{document}

\begin{flushright}
UT-02-59\\
DESY-02-192\\
hep-ph/0211115
\end{flushright}

\begin{center}
  {\large\bf Non-thermal Dark Matter from Affleck-Dine
  Baryogenesis}~\footnote{ Talk presented by K.~Hamaguchi at SUSY'02,
  DESY Hamburg, Germany, June 2002.  This talk is based on the works
  in Ref.~\cite{FujiiHama-1,FujiiHama-2}.}
  
  \vspace{0.5cm}
  
  {\bf Masaaki Fujii~$^a$ and K. Hamaguchi~$^b$}

  \vspace{0.5cm}

  $^a$ {\it Department of Physics, University of Tokyo, Tokyo 113-0033, Japan}

  $^b$ {\it Deutsches Elektronen-Synchrotron DESY, D-22603, Hamburg, Germany}
\end{center}

\begin{abstract}
In this talk we discuss the origin and nature of the dark matter in
the Affleck-Dine (AD) baryogenesis. The AD baryogenesis via most of
the flat directions predict formations of large Q-balls, and a great
number of the lightest supersymmetric particles (LSPs) are produced
nonthermally via the late-time decays of these Q-balls. In order to
avoid the overclosure of the universe by these nonthermally produced
LSPs, an LSP with a large pair-annihilation cross section, like
Higgsino- or Wino-like neutralino, is required, instead of the
standard Bino-like neutralino. This reveals new cosmologically
interesting parameter regions in various SUSY breaking models, which
have not attracted much attention so far.
\end{abstract}

\paragraph{Introduction}
The origins of the dark matter and the baryon asymmetry in the present
universe are big puzzles in particle cosmology. In the framework of
supersymmetry (SUSY), there is an ideal dark matter candidate, the
lightest SUSY particle (LSP). On the other hand, the minimal SUSY
standard model (MSSM) also provides an interesting mechanism to
generate baryon asymmetry quite effectively, by using the flat
directions in the scalar potential carrying baryon and/or lepton
number: that is, the Affleck-Dine (AD) baryogenesis~\cite{AD}. In this
talk, we point out that the Higgsino- or Wino-like neutralino
naturally becomes the dominant component of the dark matter, if either
of them is the LSP and if the AD mechanism is responsible for the
generation of the baryon asymmetry in the present universe. This
reveals new cosmologically interesting parameter regions in various
SUSY breaking models, where the most extensively studied LSP as a dark
matter candidate has been the Bino-like neutralino. We also comment on
the detection possibility of these nonthermal neutralino dark matter.

\paragraph{Affleck-Dine baryogenesis}
Let us start by briefly reviewing the AD baryogenesis~\cite{AD,DRT},
adopting the flat direction $\bar{U}\bar{D}\bar{D}$, for
example.\footnote{A classification of general MSSM flat directions and
discussions about other flat directions are available in
Ref.~\cite{FujiiHama-2}. Our main conclusion in this talk is applicable
to other flat directions as well, except for the leptonic flat
directions.} We assume that there exists the following
nonrenormalizable operator in the superpotential.\footnote{The case
without superpotential was also discussed in
Ref.~\cite{FujiiHama-2}. Actually, it has various attractive points:
there is no cosmological gravitino problem; the baryon asymmetry and
dark matter density in the present universe are determined only by the
potential of the AD field, independently of the reheating temperature
of the inflation. Recently, it has also been pointed
out~\cite{FujiiYana} that this model naturally explains the ratio of
the mass density of dark matter to that of baryons.}
\begin{eqnarray}
  W = \frac{\lambda}{M_{pl}}\bar{U}\bar{D}\bar{D}\,\,\bar{U}\bar{D}\bar{D}\,,
\end{eqnarray}
where $M_{pl}=2.4\times 10^{18}\GEV$ is the reduced Planck scale and
$\lambda$ is a coupling constant. We denote the flat direction field
by $\phi$ hereafter, and rewrite the above superpotential as follows:
\begin{eqnarray}
  W = \frac{1}{6 M^3}\phi^6\,,
\end{eqnarray}
where we have defined the effective scale $M$ including the coupling
$\lambda$. 

The relevant scalar potential for the flat direction field $\phi$ is
given by~\footnote{The thermal effects are negligible in this flat
direction as long as the reheating temperature $T_R$ of the inflation
is low enough as $T_R\lsim 10^8\GEV$~\cite{FujiiHama-2}.}
\begin{eqnarray}
  V(\phi) = (m_\phi^2 - c_H H^2)|\phi|^2 + \frac{m_{3/2}}{6 M^3}(a_m
  \phi^6 + h.c.) + \frac{1}{M^6}|\phi|^{10}\,.
  \label{scalar-potential}
\end{eqnarray}
Here, the potential terms proportional to the soft mass squared
$m_\phi^2$ and the gravitino mass $m_{3/2}$ come from the SUSY
breaking at the true vacuum. We assume that the SUSY breaking is
mediated by gravity, and take $m_{\phi}\simeq m_{3/2}|a_m|\simeq
1\TEV$.\footnote{The case of anomaly-mediated SUSY
breaking~\cite{AMSB}, where $m_{3/2}\gg m_\phi$, was also discussed in
Ref.~\cite{FujiiHama-2}.} The Hubble mass term $-c_H H^2 |\phi|^2$ is
induced by the SUSY breaking due to the finite energy density of the
inflaton~\cite{DRT}. ($H\equiv \dot{R}/R$ is the Hubble parameter, $R$
is the scale factor of the expanding universe, and the dot denotes the
derivative with cosmic time $t$.) $c_H$ is a real constant of order
unity, which depends on the couplings between the inflaton and the
$\phi$ field in the K\"ahler potential. Hereafter, we take $c_H\simeq
1$ ($>0$), which is crucial to let $\phi$ have a large expectation
value during inflation.

Now let us estimate the baryon asymmetry. The baryon number density is
given by, in terms of the AD field,
\begin{eqnarray}
  n_B = \frac{1}{3}i(\dot{\phi}^*\phi - \phi^* \dot{\phi})\,.
\end{eqnarray}
The equation of motion for the AD field is given by
\begin{eqnarray}
 \ddot{\phi} + 3 H \dot{\phi} + \frac{\partial V}{\partial \phi^*} = 0\;.
  \label{EQ-motion}
\end{eqnarray}
Then, the equation of motion for the baryon number density is written
as follows:
\begin{eqnarray}
\dot{n}_B+3 H n_B=\frac{2}{3}{\rm Im}\left(\frac{\partial
V}{\partial\phi}\phi\right) =\frac{2}{3} \frac{m_{3/2}}{M^3}\;{\rm
Im}\left(a_{m}\phi^6\right)\,,
\label{EQ-number-density}
\end{eqnarray}
By integrating this equation, we obtain the baryon number at the
cosmic time $t$ as
\begin{equation}
\left[R^{3}n_B\right](t)=\frac{2}{3} \;\frac{|a_{m}|m_{3/2}}{M^3}\int^t
\;R^3 |\phi|^6 \;{\rm sin}\theta\, dt\;,
\label{B-number}
\end{equation}
where $\theta\equiv {\rm arg}(a_{m})+6\;{\rm arg}(\phi)$.

There are three stages in the evolution of the $\phi$ field. (i)
First, during the inflation, the $\phi$ filed has a large expectation
value due to the negative Hubble mass term at the origin:
$|\phi|\simeq (H M^3)^{1/4}$. At this stage, the curvature along the
phase direction is much smaller than the Hubble parameter, and hence
the $\phi$ field has in general an arbitrary phase. Therefore, we
naturally expect that ${\rm sin}\theta={\cal O}(1)$.  

(ii) After the end of inflation, the AD field slowly rolls down toward
the origin following the gradual decrease of the Hubble parameter $H$
as $|\phi(t)|\simeq (H(t) M^3)^{1/4}\propto t^{-1/4}$. At this slow
rolling regime, the right-hand side of Eq.~(\ref{B-number}) increases
as $\propto t^{3/2}$.  Here, we have assumed the matter-dominated
universe $R\propto t^{2/3}$, which is true as long as $T_{R}\lsim
2\times 10^{10}\GEV \left(m_{\phi}/10^3\GEV\right)^{1/2}$.

(iii) Finally, the soft mass term of the AD field eventually dominates
the negative Hubble mass term at the time when $H(t_{\rm osc})\simeq
m_{\phi}$, and causes the coherent oscillation of the AD field around
the origin.  After this time, the amplitude of the AD field rapidly
decreases as $|\phi|\propto t^{-1}$, and then the production of the
baryon number terminates at the time $H_{\rm osc}\simeq m_{\phi}$.

Using the above arguments and  Eq.~(\ref{B-number}), we obtain the
baryon number density at the time $t=t_{\rm osc}$:
\begin{eqnarray}
 n_B(t_{\rm osc}) = \frac{8}{27}\delta_{\rm eff}|a_m|
  m_{3/2}\left(H_{\rm osc} M^3\right)^{1/2}
  \,,
  \label{EQ-ntosc}
\end{eqnarray}
where $\delta_{\rm eff}\equiv {\rm sin}\theta (={\cal O}(1))$. After
the completion of the reheating process of the inflation, this leads
to the following baryon asymmetry:
\begin{eqnarray}
 \frac{n_B}{s} &=& \frac{1}{4}\frac{T_R}{M_{pl}^2 H_{\rm osc}^2}n_B(t_{\rm osc})\,,
\end{eqnarray}
where $s$ is the entropy density, and $T_R$ is the reheating
temperature of the inflation.  In terms of the density parameter, it
is
\begin{eqnarray}
  \Omega_B h^2
  &\simeq&
  0.04\times \delta_{\rm eff}|a_m|
  \left(\frac{m_{3/2}}{m_\phi}\right)
  \left(\frac{1\TEV}{m_\phi}\right)^{1/2}
  \left(\frac{M}{M_{\rm pl}}\right)^{3/2}
  \left(\frac{T_R}{100\GEV}\right)\,,
\end{eqnarray}
where $h$ is the present Hubble parameter in units of $100\,\, {\rm
  km}\,\, {\rm sec}^{-1} {\rm Mpc}^{-1}$ and $\Omega_B\equiv
  \rho_B/\rho_c$ ($\rho_B$ and $\rho_c$ are the energy density of the
  baryon and the critical energy density in the present universe,
  respectively.) Here, we have used $H_{\rm osc}\simeq m_\phi$.
  Therefore, the empirical baryon asymmetry $\Omega_B h^2\simeq 0.02$
  can be explained by taking a relatively low reheating temperature
  $T_R\gsim 100\GEV$ and a reasonable set of other
  parameters. However, this is not the whole story.
\paragraph{Formation and Decay of Q-ball} 
Let us consider the epoch just after the generation of the baryon
asymmetry finishes, i.e., just after the $\phi$ field starts its
coherent oscillation. If we take into account the one-loop correction,
the mass term of the flat direction field becomes
\begin{eqnarray}
 V_{\rm mass}(\phi) = m_\phi^2
  \left[1 + K \log \left(\frac{|\phi|^2}{M_G^2}\right)\right]
  |\phi|^2\;,
\label{potential-Q-ball-formation}
\end{eqnarray}
where $M_G$ is the renormalization scale at which the soft mass
$m_\phi$ is defined, and the $K\log(|\phi|^2)$ term represents the
one-loop correction, which mainly comes from the gaugino loops. $K$ is
estimated in the range from $-0.01$ to
$-0.1$~\cite{Enq-McD-PLB425,Enq-McD-NPB538,Enq-Jok-McD}. This
potential, flatter than $\phi^2$, causes spatial instabilities of the
homogeneous $\phi$'s oscillation~\cite{KS}, and the oscillation of the
$\phi$ field fragments into inhomogeneous lumps. These lumps
eventually form non-topological solitons, ``Q-balls''~\cite{Qball}.
We should emphasize here the fact that almost all the baryon asymmetry
generated by the AD baryogenesis is absorbed in the
Q-balls~\cite{Kasu-Kawa-PRD62}. Hence, the baryon asymmetry in the
present universe must be provided by the decays of these Q-balls.

The decay temperature of the Q-ball is given by~\cite{Qballdecay},
\begin{eqnarray}
  T_d \lsim 2\GEV\times 
  \left(\frac{0.03}{-K}\right)^{1/2}
  \left(\frac{m_\phi}{1\TEV}\right)^{1/2}
  \left(\frac{10^{20}}{Q_i}\right)^{1/2}\,,
  \label{TdQ}
\end{eqnarray}
where $Q_i$ is the initial charge of the individual Q-ball. In the
case of flat directions with the scalar potential
(\ref{scalar-potential}), it turns out to be~\cite{FujiiHama-2}
\begin{eqnarray}
  Q_i\simeq 10^{20}\times 
  \delta_{\rm eff}|a_m|
  \left(\frac{m_{3/2}}{m_\phi}\right)
  \left(\frac{1\TEV}{m_\phi}\right)^{3/2}
  \left(\frac{M}{M_{\rm pl}}\right)^{3/2}
  \,.
\end{eqnarray}
Therefore, the Q-ball decay occurs below $\sim$(a
few)$\GEV$. Actually, in most of the flat directions, the charge of
the formed Q-ball is as large as $Q_i\sim
10^{19}$--$10^{26}$~\cite{FujiiHama-2}, which is basically determined
by dimensions of the non-renormalizable operator that lifts the
relevant flat directions. Hence the decay temperature $T_d$ of the
Q-ball varies in the range of $T_d\sim 1\MEV$--(a few)$\GEV$.

\paragraph{Nonthermal production of LSP}
Because the Q-ball consists of squarks, its decay produces quarks and
supersymmetric particles. The quarks become baryon asymmetry, which is
requisite to the big-bang nucleosynthesis, while the supersymmetric
particles eventually decay into LSPs. Thus, the number density of the
LSPs, $n_{\rm LSP}$, is related to the baryon asymmetry:
\begin{eqnarray}
  n_{\rm LSP} \ge n_\phi \ge 3 n_B\,,
  \label{nLSPnB}
\end{eqnarray}
where $n_\phi$ is the number density of the $\phi$ field ($=$ squarks)
in the Q-balls.  An important point here is that the Q-ball decay
occurs {\it below} the freeze-out temperature of the LSPs, which is
typically given by $T_f\sim m_{\rm LSP}/20$. ($m_{\rm LSP}$ is the
mass of the LSP.)  Namely, the LSPs are never thermalized after they
are produced by the Q-ball decay. If there is no pair annihilation of
LSPs afterwards, therefore, the relation in Eq.~(\ref{nLSPnB})
maintains until present, which results in~\cite{Enq-McD-NPB538}:
\begin{eqnarray}
  \Omega_{\rm LSP}\ge 3\frac{m_{\rm LSP}}{m_p}\Omega_B
  = 12 
  \left(\frac{m_{\rm LSP}}{100\GEV}\right)
  \left(\frac{\Omega_B}{0.04}\right)
  \,,
\end{eqnarray}
where $m_p$ is the nucleon mass. This conflicts
with the observed dark matter density $\Omega_{DM}\simeq 0.3$, unless
the LSP mass is extremely small:
\begin{eqnarray}
  m_{\rm LSP}\le
  3\GEV\left(\frac{\Omega_{\rm LSP}}{0.3}\right)\left(\frac{0.04}{\Omega_B}\right)
  \,.
\end{eqnarray}
Consequently, the formation and late time decays of the Q-balls is a
serious obstacle for the AD baryogenesis.  This is actually the case
in the standard parameter regions where the thermal relics of
Bino-like LSPs lead to a cosmologically interesting mass density of
dark matter.

Here, we consider a simple solution to this
problem~\cite{FujiiHama-1}: an LSP with a large pair-annihilation
cross section, like Higgsino- or Wino-like neutralino. If significant
pair-annihilations of LSPs occur after they are produced, the relation
in Eq.~(\ref{nLSPnB}) no longer holds. The final abundance of the LSP
is obtained by solving the Boltzmann equation analytically, which
leads to~\cite{FujiiHama-1,FujiiHama-2}
\begin{eqnarray}
 Y_{\rm LSP}(T)
  \simeq
  \left[
   \frac{1}{Y_{\rm LSP}(T_d)}
   +
  \sqrt{
   \frac{8\pi^2 g_*(T_d)}{45}
   }
   \vev{\sigma v}
    M_{\rm pl}
    (T_d - T)
   \right]^{-1}
   \,,
   \label{EQ-analytic-YT}
\end{eqnarray}
where $Y_{\rm LSP}\equiv n_{\rm LSP}/s$, $g_*(T)$ is the effective
degrees of freedom at temperature $T$, and $\vev{\sigma v}$ is the the
thermally averaged annihilation cross section of the LSP. If initial
abundance $Y_{\rm LSP}(T_d)$ is large enough, the final abundance
$Y_{\rm LSP}^0$ for $T \ll T_d$ is given by
\begin{eqnarray}
 Y_{\rm LSP}^0
  \simeq Y_{\rm LSP}^{\rm approx}\equiv
  \left[
   \sqrt{
   \frac{8\pi^2 g_*(T_d)}{45}
   }
   \vev{\sigma v}
   M_{\rm pl}
   T_d
   \right]^{-1}
   \,.
   \label{EQ-Ychi-analytic}
\end{eqnarray}
Therefore, in this case, the final abundance $Y_{\rm LSP}^0$ is determined
only by the Q-ball decay temperature $T_d$ and the annihilation cross
section of the LSP $\vev{\sigma v}$, independently of the initial
value $Y_{\rm LSP}(T_d)$ as long as $Y_{\rm LSP}(T_d)\gg Y_{\rm LSP}^{\rm
approx}$.  In terms of the density parameter $\Omega_{\rm LSP}$, it is
rewritten as
\begin{eqnarray}
 \Omega_{\rm LSP}
 \simeq
  0.5
  \left(\frac{0.7}{h}\right)^2
  \times
  \left(
   \frac{m_{\rm LSP}}{100 \GEV}
   \right)
   \left(
    \frac{10^{-7}\GEV^2}{\vev{\sigma v}}
    \right)
    \times
    \left(
     \frac{100 \MEV}{T_d}
     \right)
     \left(
      \frac{10}{g_*(T_d)}
      \right)^{1/2}
      \,,
      \label{Omega-ana}
\end{eqnarray}
Therefore, the LSP needs a pair annihilation cross section as large as
$\vev{\sigma v}\sim 10^{-8}\mbox{--}10^{-6}\GEV^{-2}$ so as to obtain
the correct mass density as a dominant component of dark matter for
the typical decay temperature of Q-balls.  Interestingly, Higgsino-
and Wino-like LSPs have pair annihilation cross sections just in the
desired range, and naturally leads to the correct mass density of dark
matter.

\begin{figure}[t!]
 \centerline{\psfig{figure=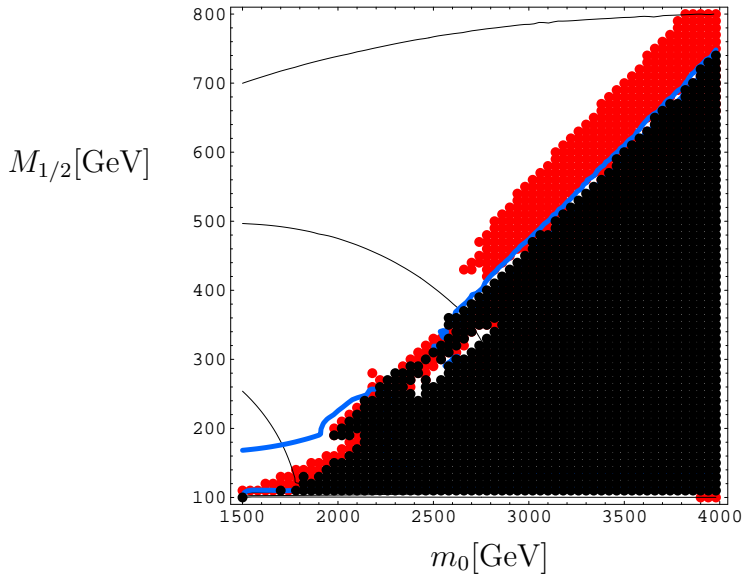,height=7cm}}
 \begin{picture}(0,0)
  \put(40,150){$M_{1/2}$[GeV]}  
   \put(200,1){$m_{0}$[GeV]}
\normalsize
 \end{picture}
\vspace{0.5cm}
 \caption{The allowed region in the mSUGRA scenario with ${\rm
tan}\beta=15$ and $A_{0}=0$ in the $(m_{0}$--$M_{1/2})$ plane.  In the
red shaded region, non-thermally produced LSPs via decays of Q-balls
result in $0.05 \le \Omega_{\rm LSP} h^2 \le 0.5$ for $1\MEV \le T_d
\le 10\GEV$. The black shaded region is where the electroweak symmetry
breaking cannot be implemented. The region below the blue (thick) line
is excluded by the chargino mass bound $m_{\chi^{\pm}}\gsim 105\GEV$.
The contours of the light Higgs boson mass are given by the black
(thin) lines, which correspond to $m_{h}=117,\;120,\;122\GEV$,
respectively. {}From Ref.~\cite{FujiiHama-2}.}
\label{fig_1}
\end{figure}

\paragraph{Parameter space and possibility of detections}
We have studied in detail the parameter space where the nonthermal LSP
from the Q-ball decay becomes the dominant component of the dark
matter, adopting several SUSY breaking models. The Wino-like
neutralino dark matter is realized in a wide parameter regions in the
anomaly-mediated SUSY breaking models~\cite{AMSB} and the no-scale
models with nonuniversal gaugino masses~\cite{nonUniv}. Higgsino dark
matter is realized in the minimal supergravity scenario, in the
so-called ``focus point'' region~\cite{focuspoint}. Here, we show an
example of the latter case in Fig.~\ref{fig_1}. We have also
investigated direct and indirect detection of the neutralino dark
matter in these regions, and found that there is an intriguing
possibility to detect them in various next generation dark matter
search experiments~\cite{FujiiHama-2}. Actually, the possibility of
direct and indirect detection of them is much larger than that of the
standard Bino-like neutralino.

\paragraph{Summary} 
The formation and late time decays of Q-balls are inevitable
consequences of the Affleck-Dine baryogenesis via most of the flat
directions. In order to avoid the overclosure of the universe by the
nonthermal LSPs, which are generated from Q-ball decay, the LSP should
have a large pair annihilation cross section. The possible candidates
are Higgsino- and Wino-like neutralinos.

We emphasize that, if the Higgsino- or Wino-like neutralino dark
matter is indeed detected at future experiments, in the parameter
regions we have shown, then it suggests the existence of nonthermal
source of these LSPs, since thermal abundance of them would be too
small to be the dominant component of the dark matter. The Q-balls
produced via the Affleck-Dine baryogenesis is the most promising
candidate of such a nonthermal source of the LSPs.



\begin{thebibliography}{99}
\bibitem{FujiiHama-1} M.~Fujii and K.~Hamaguchi, 
Phys.\ Lett.\ B {\bf 525} (2002)
  143.  

\bibitem{FujiiHama-2}
  M.~Fujii and K.~Hamaguchi,
  Phys.\ Rev.\ D {\bf 66} (2002) 083501.

\bibitem{AD}
  I.~Affleck and M.~Dine,
  Nucl.\ Phys.\ B {\bf 249} (1985) 361.

\bibitem{DRT}
  M.~Dine, L.~Randall and S.~Thomas,
  Phys.\ Rev.\ Lett.\  {\bf 75} (1995) 398
  and
  Nucl.\ Phys.\ B {\bf 458} (1996) 291.

\bibitem{FujiiYana}
  M.~Fujii and T.~Yanagida,
  Phys.\ Lett.\ B {\bf 542} (2002) 80.

\bibitem{AMSB}
  L.~Randall and R.~Sundrum,
  Nucl.\ Phys.\ B {\bf 557} (1999) 79;
  G.~F.~Giudice, M.~A.~Luty, H.~Murayama and R.~Rattazzi,
  JHEP {\bf 9812} (1998) 027;
  J.~A.~Bagger, T.~Moroi and E.~Poppitz,
  JHEP {\bf 0004} (2000) 009.

\bibitem{Enq-McD-PLB425}
  K.~Enqvist and J.~McDonald,
  Phys.\ Lett.\ B {\bf 425} (1998) 309.

\bibitem{Enq-McD-NPB538}
  K.~Enqvist and J.~McDonald,
  Nucl.\ Phys.\ B {\bf 538} (1999) 321.

\bibitem{Enq-Jok-McD}
  K.~Enqvist, A.~Jokinen and J.~McDonald,
  Phys.\ Lett.\ B {\bf 483} (2000) 191.


\bibitem{KS}
  A.~Kusenko and M.~E.~Shaposhnikov,
  Phys.\ Lett.\ B {\bf 418} (1998) 46.




\bibitem{Qball}
  S.~R.~Coleman,
  Nucl.\ Phys.\ B {\bf 262} (1985) 263
  [Erratum, ibid.\ B {\bf 269} (1985) 744].

\bibitem{Kasu-Kawa-PRD62}
  S.~Kasuya and M.~Kawasaki,
  Phys.\ Rev.\ D {\bf 62} (2000) 023512.

\bibitem{Qballdecay}
  A.~G.~Cohen, S.~R.~Coleman, H.~Georgi and A.~Manohar,
  Nucl.\ Phys.\ B {\bf 272} (1986) 301.


\bibitem{nonUniv}
  S.~Komine and M.~Yamaguchi,
  Phys.\ Rev.\ D {\bf 63} (2001) 035005.


\bibitem{focuspoint}
J.~L.~Feng, K.~T.~Matchev and T.~Moroi,
Phys.\ Rev.\ Lett.\  {\bf 84} (2000)  2322;
Phys.\ Rev.\ D {\bf 61}  (2000) 075005.



\end{thebibliography}
\end{document}